\documentstyle[multicol,aps,prb,epsfig]{revtex}
\begin{document}
\baselineskip=0.5cm
\renewcommand{\thefigure}{\arabic{figure}}
\title{Hard-core Yukawa model for two-dimensional charge stabilized
colloids}
\author{R. Asgari$^a$, B. Davoudi$^{a,b}$ and B. Tanatar$^c$}
\address{
$^a$(IPM) Institute for Studies in Theoretical Physics and
    Mathematics, Tehran, P.O.Box 19395-5531, Iran\\
$^b$Scuola Normale Superiore, Piazza dei Cavalieri 7, 56126 Pisa, 
Italy\\ 
$^c$Department of Physics, Bilkent University, Bilkent, Ankara}
\date{\today}
\maketitle
\begin{abstract}
The hyper-netted chain (HNC) and Percus-Yevick (PY) approximations 
are used to study the phase diagram of a simple hard-core Yukawa model
of charge-stabilized colloidal particles in a two-dimensional 
system. We calculate the static structure factor and 
the pair distribution function over 
a wide range of parameters. Using the statics correlation
functions we present an estimate for the liquid-solid phase 
diagram for the wide range of the parameters.
\end{abstract}
\begin{multicols}{2}

\section{Introduction}

Systems composed of small particles with sizes of 10's of
nanometers dispersed in solvents are known as colloidal
suspensions.\cite{ptoday} With increasing interest in the properties 
of complex fluids and their various thermodynamic phases, colloidal 
systems are a subject of vast experimental and theoretical
investigations.\cite{pieranski1,safran} 

The two-dimensional (2D) colloids are of special interest for
several reasons. First, colloidal systems can be realized rather
easily experimentally, as in air-water interfaces or between two
parallel glass plates, and advances in measurement techniques such 
as digital video microscopy bring about a wealth of information 
on these systems.\cite{Pieranski,tang,murray,Marcus1,Marcus2} 
Second, the nature of melting of 2D crystals is 
different than their 3D counterparts, making them a subject of intense
theoretical and experimental investigations.\cite{Weber,blandon,terao} 
Colloidal suspensions in the fluid phase may crystallize as 
their density increases or the range of interaction changes. 
Existence of a glassy phase\cite{crys2} and re-entrant 
melting\cite{wei,bubeck} are examples of various motivating 
derives behind some of the recent work. Possibilities of photonic 
band-gap structures made out of colloidal crystals promise interesting 
applications.\cite{john} 

In this work, we first use the HNC and PY approximations for a 
comparison between the two approaches for the pure hard-core 
potential (PHCP) as a model for a colloidal system. It is well 
known that the simple HNC approximation, 
which omits the elementary diagrams, gives a good description of
the large interparticle distance behavior. On the other hand, the 
PY approximation yields better results for the short interparticle 
distances.\cite{Rosenfeld,Rogers,rzysko} 
In the next step, we employ the widely used hypernetted-chain (HNC)
approximation to study the liquid state correlation functions
and freezing transition of charge-stabilized colloidal particles
in two-dimensions interacting via the hard-core Yukawa
potential (HCYP). Our basic aim is to test how well the HNC method
models the static properties of a colloidal system. Similar
calculations\cite{Davoudi} in 3D resulted in good agreement between the
predictions of HNC theory and Monte Carlo simulations. To this end,
we calculate the pair-correlation function and the static structure
factor for a range of densities and temperatures. We discuss our
results in comparison to other theoretical approaches and available
experiments. 

\section{Model and Theory}

\subsection{Hard-Core Yukawa Model}

In this section, we introduce the model of two-dimensional 
charged colloidal particles. We consider the colloidal particles
trapped in a surface energy well at an air-water interface modeled 
by a hard-core and a Yukawa interaction.\cite{Pieranski} 
The pair potential $V(r)$ between the colloidal particles is
given by
\begin{equation}\label{HCYP}
V(r)=\left\{ 
\begin{array}{ll}
\infty  & \mbox{for $r<\sigma$\, ,}\\
  \lambda \frac{\exp{[-\kappa (r-\sigma)]}}{r}& \mbox{for $r>\sigma$\, ,}
\end{array}
\right.
\end{equation}
where $\lambda$ is the strength of the interaction, $\sigma$ is
the hard-core radius, and $\kappa$ is the inverse of the Debye 
screening length of the colloidal system. PHCP without the
Yukawa tail is obtained by setting $\lambda=0$. The form of the
interaction potential between charge-stabilized colloids is
different from that between charged colloidal particles
as inferred from the recent experiments of Marcus and
Rice.\cite{Marcus1,Marcus2} The role of long-range forces in
colloidal systems has been emphasized by Noro {\it et
al}.\cite{noro}

\subsection{HNC and PY formalism}

Given the above potential to describe interactions between charged
colloidal particles, the evaluation of correlation functions reduces 
to a problem in classical liquid state theory\cite{Hansen}. One of 
the basic quantities of interest is the pair-distribution function 
defined by
\begin{equation}\label{PDF}
g(|\bf r-\bf r'|)=\left\langle\sum_{i=1}^N\sum_{j\neq i}^N
\delta(\bf r-\bf r_i)\delta(\bf r'-\bf r_j)\right\rangle/\rho^2,
\end{equation}
where $\rho$ is the density of the liquid. The pair-distribution
function gives the probability of finding a particle at $r'$ when we 
fix another particle at $r$. Other relevant quantities are the 
pair-correlation function $h(r)=g(r)-1$, and the static structure 
factor defined by
\begin{equation}\label{SSF}
S(q)=1+\rho\int d{\bf r} h(r)\exp\left(i{\bf q}.{\bf r}\right)\, .
\end{equation}
The pair-correlation function for a simple, isotropic 
fluid can be decomposed using the Ornstein-Zernike 
relation\cite{Hansen,Menon} 
\begin{equation}\label{OZE}
h(r)=C(r)+\rho\int d^2 r'\,C(|\bf r-\bf r'|)h(\bf r')\, ,
\end{equation}
where $C(r)$ is called the direct correlation function
and can be related to the structure factor through
\begin{equation}\label{SC}
S(q)=1/[1-\rho C(q)]\, .
\end{equation}
A closure relation between $h(r)$ and $C(r)$ is needed to
supplement the Ornstein-Zernike relation. There are some closure
relations between $h(r)$ and $C(r)$. In the following we use the 
well-known the HNC and PY approximations. In the HNC approximation, 
the closure relation takes the form\cite{Hansen,Menon}
\begin{equation}\label{SOZE}
C(r)=\exp\left[-\beta V(r)+Y(r)\right]-1-Y(r)\, ,
\end{equation}
where $V(r)$ is the interparticle pair potential and by
construction $Y(r)=h(r)-C(r)$. $\beta=1/(k_BT)$ is the inverse
temperature. On another hand, the PY closure 
relation is given by\cite{Hansen}
\begin{equation}\label{PY}
C(r)=\left(e^{-\beta V(r)}-1\right)\left(Y(r)+1\right)\, ,
\end{equation}
The self-consistent solution of
Eqs.~(\ref{OZE}), ~(\ref{SOZE}) or ~(\ref{PY}) gives the 
information on the correlations in the liquid state
within the HNC or PY schemes.

\section{Numerical results and Discussion}

In this section, we present the results of our numerical 
calculation of the HNC and PY equations for a system of colloidal
particles interacting with HCYP. 
We have solved Eqs.\,~(\ref{OZE}),~(\ref{SOZE}) or Eq.\,~(\ref{PY}) 
for HNC and PY approximations, respectively. One can find solutions 
for these set of equations by numerical iteration taking the potential 
defined in Eq.~(\ref{HCYP}) as input. 
We have obtained the correlation
functions for a wide range of values of the density $\rho^*=\sigma^2 
\rho$, screening length $\kappa\sigma$, and inverse temperature
$\beta \lambda$. 
More specifically, we have calculated the pair-correlation function, 
the static structure factor, the collective mode energy, and direct 
correlation function which is related to the effective potential. 
We have then used the Hansen-Verlet\cite{HansenVerlat} criterion to 
estimate the phase line for liquid-solid transition.
 
\begin{figure}
\centerline{\mbox{\psfig{figure=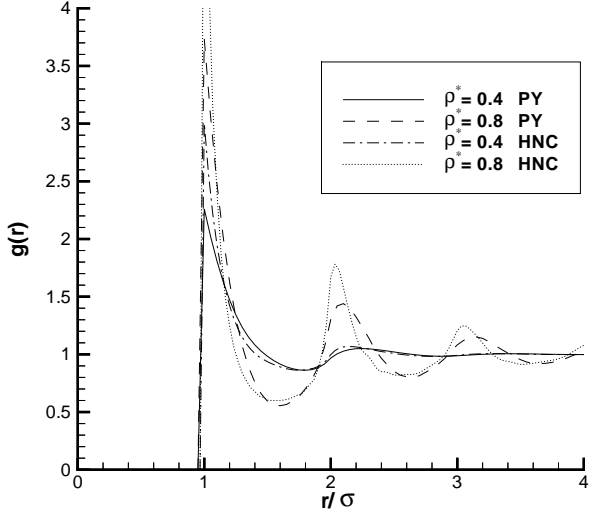, angle =0, width =9 cm}}} 
\caption{The pair-distribution function for PHCP in 
terms of $k\sigma$ at $\rho=0.4\, {\rm and }\, 0.8$ within the HNC 
and PY approximations.}
\label{Fig1}
\end{figure}

\begin{figure}
\centerline{\mbox{\psfig{figure=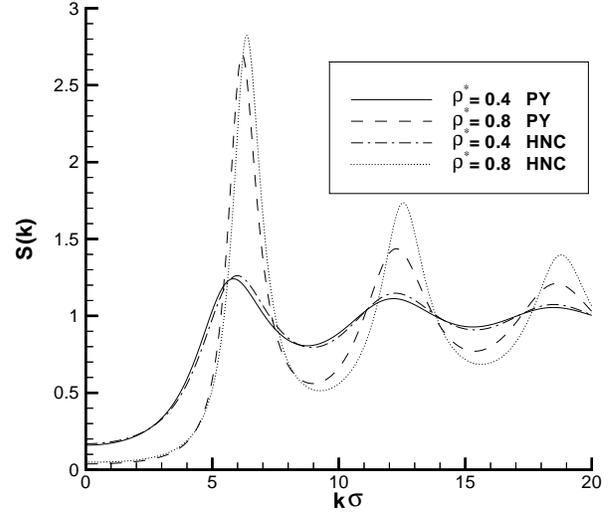, angle =0, width =9 cm}}} 
\caption{The static structure factor for PHCP in 
terms of $k\sigma$ at $\rho=0.4\, {\rm and }\, 0.8$ within the HNC 
and PY approximations.}
\label{Fig2}
\end{figure}

As a comparison, firstly, we compare the results between the HNC and 
PY approximations. To show the differences, we compare the results 
for two extreme regions, low and high densities. In Fig.~\ref{Fig1}, 
we show the pair-distribution function at 
$\rho^*=0.4\,\, {\rm and}\,\,0.8$ within the HNC and PY approximations 
for PHCP. One can see that the differences between two approaches 
increase by increasing density. It is evident that we have a sharp 
peak in $r=\sigma$ and the height of the peak increases as we increase 
the density. This means that the probability of finding a 
particle at $r=\sigma$ when we fix a particle in the origin increases 
with increasing density. One can observe the long-range 
oscillatory behavior in pair-distribution function which is a sign for 
liquid-solid phase transition.
Using Eq.\,(\ref{SSF}), one can find the static structure factor. 
We depict in Fig.~\ref{Fig2} the static structure factor at 
$\rho^*=0.4\,\, {\rm and}\,\,0.8$ within the HNC and PY approximations 
for PHCP. There is a peak around $k=2\pi/\sigma$ which increases with 
increasing density. A sharp peak in the structure factor at 
$k=2\pi/\sigma$ produces long-range oscillatory behavior with 
periodicity $\sigma$ in the pair-distribution 
function which is a signature for the long-range order or the solid 
phase. Furthermore, the height of this peak is very important in 
predicting of the liquid-solid phase transition. The Hansen-Verlet 
criterion simply states that the phase transition takes place when the 
value of the first peak in $S(k)$ reaches a special 
constant. It is interesting that the difference between the value of 
the static structure factor within the two approximations near the 
first peak is small. This small difference allows the use of HNC 
results to be reasonable to predict the liquid-solid phase transition 
for PHCP. We used the recent results of Weber, Marx and 
Binder \cite{Weber} 
for the density at which the phase transition takes place to determine 
the height of this peak in our results. This value in our results 
is about $S_{max}\cong3.5$, corresponding to $\rho^*=0.89$ within the HNC approximation.

\begin{figure}
\centerline{\mbox{\psfig{figure=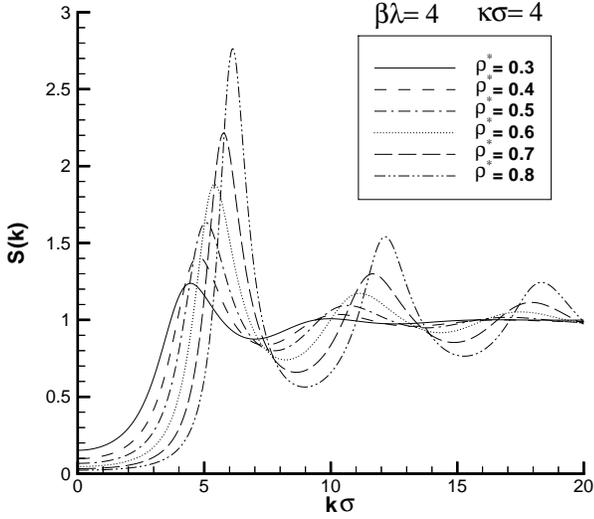, angle =0, width =9 cm}}} 
\caption{The static structure factor in terms of $k\sigma$
at $\beta\lambda=4$, $\kappa\sigma=4$ and different values of 
$\rho$ .}
\label{Fig3}
\end{figure}

We now report the results for HCYP within the HNC approximation. 
The static structure factor is shown in Fig.\,\ref{Fig3} 
at fixed $\beta\lambda=4$ and $\kappa\sigma=4$ for different 
values of $\rho^*$. We observe that with increasing density 
the peak structure in the static structure factor 
increases. Again, the increasing peak height of static 
structure factor shows the tendency toward the formation of 
solid phase. We also can observe that there is a shift in the 
position of the peak, which means in the high density regime the 
hard-core part of the potential is dominant compared to the 
Yukawa part and the position of the peak shifts to the position
for the hard-core one.

\begin{figure}
\centerline{\mbox{\psfig{figure=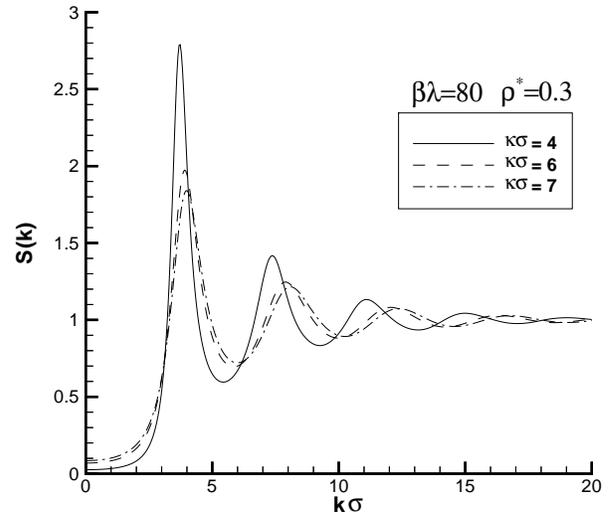, angle =0, width =9 cm}}} 
\caption{The static structure factor in terms of $k\sigma$
at $\rho^*=0.3$, $\beta\lambda=80$ and different values of 
$\kappa\sigma$.}
\label{Fig4}
\end{figure}

\begin{figure}
\centerline{\mbox{\psfig{figure=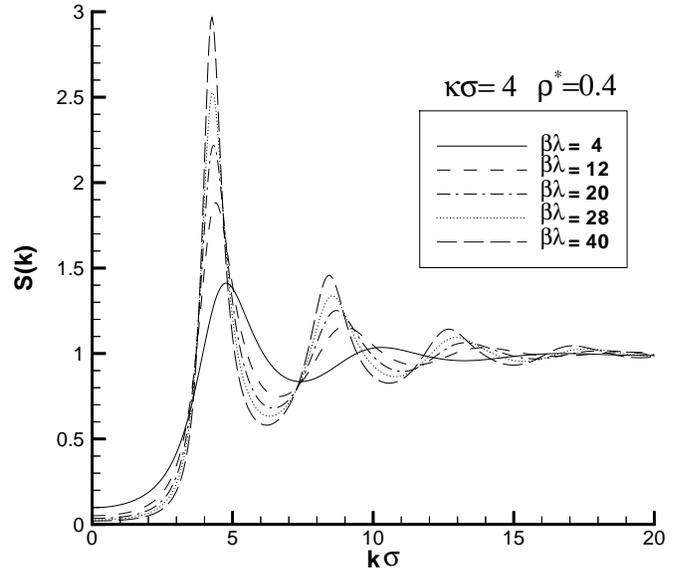, angle =0, width =10 cm}}} 
\caption{The static structure factor in terms of $k\sigma$
at $\rho^*=0.4$, $\beta\lambda=4$ and different values 
$\kappa\sigma$.}
\label{Fig5}
\end{figure}

We observe the same behavior by fixing $\kappa\sigma$ and
$\rho^*$, and changing $\beta\lambda$ or by fixing $\beta\lambda$ 
and $\rho^*$, and changing $\kappa\sigma$. The above argument means
that by changing the strength and range of the Yukawa part or the 
temperature, one can reach the phase transition point. 
The results at $\rho^*=0.3$, $\beta\lambda=80$ for different 
values of $\kappa\sigma$ and at $\rho^*=0.4$, $\beta\lambda=4$ 
for different values $\kappa\sigma$ are depicted in 
Figs.\,\ref{Fig4} and \ref{Fig5}. 

It is evident from Fig.\,\ref{Fig4} that
the position of the peak is different from the high density limit, 
which implies, for low density limit the Yukawa part has the major 
contribution in the liquid-solid phase transition. The position of 
the peak also has some information about periodicity in 
pair-distribution function. Generally, the position of the peak near 
the phase transition point for HCYP only
depends on the value of the density and it is insensitive to other 
parameters.

\begin{figure}
\centerline{\mbox{\psfig{figure=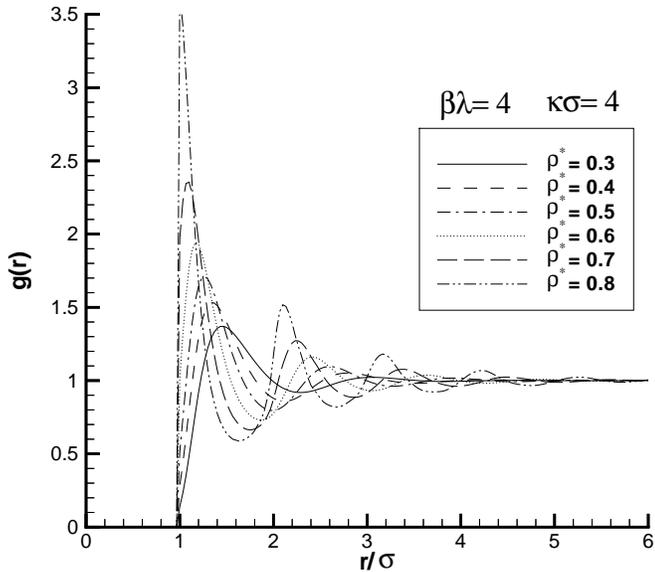, angle =0, width =10 cm}}} 
\caption{The pair distribution function in terms of $r/\sigma$
at $\beta\lambda=4$ and $\kappa\sigma=4$ and different value 
of $\rho^*$.}
\label{Fig6}
\end{figure}

In Fig.\,\ref{Fig6} we show the pair-distribution function at 
fixed $\beta\lambda=4$, $\kappa\sigma=4$ and different values of 
$\rho^*$. It is interesting to note that we have a shift toward 
$r=\sigma$ as we increase the density, which again means that the 
hard-core part has major contribution in the high density limit. It 
is clear that the period of the oscillations also changes as the 
density increases.

\begin{figure}
\centerline{\mbox{\psfig{figure=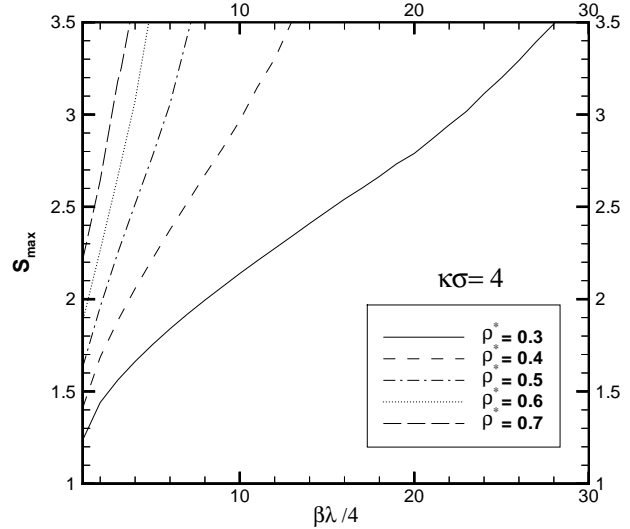, angle =0, width =9 cm}}} 
\caption{The value of the first peak in the static structure 
factor in term of $\beta\lambda$ at $\kappa\sigma=4$ and 
$\rho=0.4, 0.5, 0.6, 0.7$.}
\label{Fig7}
\end{figure}

It is possible to study the evolution of the first peak in the static
structure factor when the physical quantities change, to extract 
some useful information about liquid-solid phase transition. We show 
in Fig.\,\ref{Fig7}, the variation of this peak in terms of 
$\beta\lambda$ for a set of densities 
$\rho^*=0.3,\,0.4,\,0.5,\,0.6,\,0.7$ at $\kappa\sigma=4$. 
It is clear from the figure that the Yukawa part is important 
only for low density and it is the source of liquid-solid phase 
transition in the low density region. It is interesting that 
$\beta\lambda$ at which the phase transition takes place also shifts 
to the right very rapidly with decreasing density, because 
the mean distance between two particles increases with reducing 
density and the effect of the Yukawa interaction 
decreases exponentially as the mean distance increases.

\begin{figure}
\centerline{\mbox{\psfig{figure=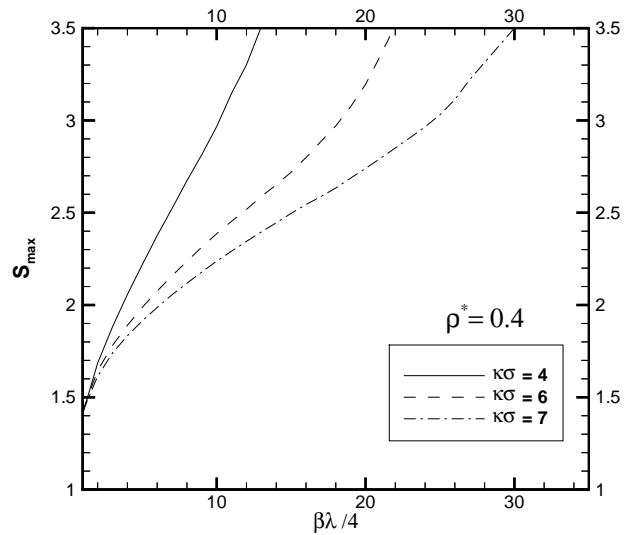, angle =0, width =9 cm}}} 
\caption{The value of the first peak in the static structure 
factor in term of $\beta\lambda$ at $\rho=0.4$ and 
$\kappa\sigma=4,\,6,\,7$.}
\label{Fig8}
\end{figure}

In Fig.\,\ref{Fig8} we show the variation of the first peak in terms 
of $\beta\lambda$ at $\rho=0.4$ and $\kappa\sigma=4,\,6,\,7$. It is 
observed that with increasing $\kappa\sigma$, the curves $S_{\rm
max}$ are going to tend to a constant, which means that hard-core 
part again has the major effect on this system in this region and 
we can thus observe the phase transition with changing density.

\begin{figure}
\centerline{\mbox{\psfig{figure=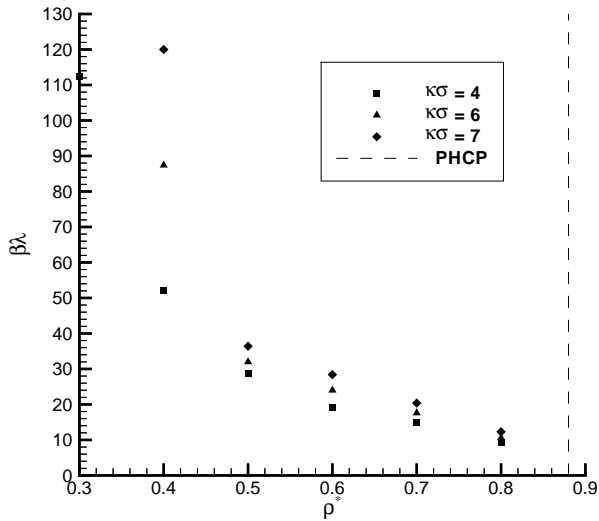, angle =0, width =9 cm}}} 
\caption{The liquid-solid phase line transition at 
$\kappa\sigma=4,\,6,\,7 $. The dashed line represents the 
liquid-solid phase transition line for the PHCP.}
\label{Fig9}
\end{figure}

From the previous figures and using the Hansen-Verlet criterion,
we can obtain the liquid-solid phase transition line.  We depict 
in Fig.\,\ref{Fig9} the phase line for different values of 
$\kappa\sigma$. It is interesting to note that all 
phase separation lines tend to the same point in the high density 
limit, which is reasonable.
Another important point is that with increasing $\kappa\sigma$ 
the phase line tends to the line $\rho^*=0.89$. This line is 
equivalent to the liquid-solid phase transition line for pure 
hard-core potential. 

The phase diagram of colloidal systems in 2D is of particular
recent interest because of the subtleties in freezing or melting
transition. Bladon and Frankel\cite{blandon} using an attractive
aquare-well potential found a first order
melting transition where the system goes into a liquid phase
from a crystal phase. However, molecular dynamics\cite{zangi} 
simulations and recent experiments\cite{pallop} show solid-to-hexatic 
and hexatic-to-liquid phase transitions if there is an attractive
part to the colloidal potential. In an earlier work 
L{\"o}wen\cite{lowen} did not find a hexatic phase in a Yukawa 
system. It appears that how the form of
the interaction potential influences the possible phases in 2D
colloidal systems is not entirely clear. Issues such as
attractive or repulsive nature, hard or soft core, and long
range part of the potential remain to be systematically
investigated. In our analysis, we used the Hansen-Verlet
criterion to determine the liquid-solid freezing transition
which seem consistent with available results.

In summary, we have used in this paper the HNC theory to study the correlation
functions of a 2D hard-core Yukawa fluid as a model for
colloidal systems. We have investigated the effect of the Yukawa
tail in the model potential by comparing our results with those
for purely hard-core potential. The reliability of our
correlation functions at small distances are tested against the
PY approximation. Using the Hansen-Verlet criterion which uses
the value of the first peak in the static structure factor, we
obtained the phase diagram showing the liquid-solid transition.

\acknowledgements
{This work was partially supported by the Scientific and Technical 
Research Council of Turkey (TUBITAK) under Grant No. TBAG-2005, by 
NATO under Grant No.  SfP971970, and by the Turkish Department of 
Defense under Grant No. KOBRA-001. One of us  (B. D.) acknowledges because of a post-doctral grant by Scula Normale Superiore under the MURST PRIN1999 Initiative. We would like to thank Prof. M. P. Tosi and Dr. A. L. Demirel for useful discussions and comments.}

\end{multicols}
\end{document}